\documentclass[aps,prl,preprint]{revtex4}
\usepackage{amsfonts}
\usepackage{amsmath}
\usepackage{amssymb}
\usepackage{graphicx}

\begin{document}

\title{ {\it Ab initio} determination of ion-traps in silver-doped chalcogenide glass}

\author{I. Chaudhuri}
\affiliation{Department of Physics and Astronomy, Ohio University, Athens OH 45701, USA}
\author{F. Inam}
\affiliation{Department of Physics and Astronomy, Ohio University, Athens OH 45701, USA}
\author{D. A. Drabold}
\affiliation{Department of Physics and Astronomy, Ohio University, Athens OH 45701, USA}
\affiliation{Department of Chemistry, University of Cambridge, Lensfield Road, Cambridge, CB2 1EW, United Kingdom}

\date{\today}

\begin{abstract}
We present a microscopic picture of silver dynamics in GeSe$_{3}$:Ag glass obtained from {\it ab initio} simulation. The dynamics of Ag is explored at two temperatures, 300K and 700K. In the relaxed network, Ag occupies bond centers between suitably separated host sites.   At 700K, Ag motion proceeds via  a trapping-release dynamics, between ``super traps" or cages consisting of multiple bond-center sites in a small volume. Our work offers a first principles identification of trapping centers invoked in current theories,  with a description of their properties. We compute the charge state of the Ag in the network, and show that it is neutral if weakly bonded and Ag$^+$ if in a trapping center.
\end{abstract}


\maketitle

One of the outstanding problems of solid-state ionics is the nature of the hopping of metal atoms in fast-ion conductors. As a step toward  a general atomistic picture, we offer a detailed {\it ab initio} study of Ag dynamics in a glassy germanium selenide matrix. In agreement with a previous study\cite{tafen242}, we show that trapping centers exist, and for the first time describe their atomistic nature in detail. Our work is also related to the theory of batteries, the prime technological application of solid electrolytes. Finally, the promising ``Programmable Metallization Cell", a novel non-volatile computer memory device \cite{kozicki06} is made from the materials we describe here. 

Silver-doped Ge-Se glasses exhibit a fascinating range of phenomena: among them high ionic conductivity\cite{kawasaki99} and photo-doping\cite{kolobov91}, connected with high Ag diffusion\cite{bychkov96,urena05}. An empirical model of Elliott \cite{elliott94}, provides an estimate of the activation energies of ionic conductivities. In this model the Ag ions are assumed to be surrounded by cations (Se), and Coulombic and polarization interactions give rise to the potential barriers, leading to the hopping dynamics of Ag. The first MD simulations, using an empirical potential, were due to Iyetomi {\it et. al}\cite{iyetomi00}, who detected phase separation of the Ag atoms in GeSeAg glasses. Recently, we have reported the existence of Scher-Lax-Phillips traps\cite{phillips96},  in {\it ab initio} models of (GeSe$_{3})_{1-x}Ag_{x}$ (for $x=0.10$ and $x=0.15$)\cite{tafen242}. Our calculations are related to a recently-published percolative free-volume model\cite{adams}, and provide microscopic identification of the trapping centers proposed by these and other authors. 

In this paper, we find that Ag atoms sit preferentially near the midpoint between host (Ge or Se) atoms separated by about 5.0\AA ~ at T=0, and we name these energy minima ``Trapping Centers" or TC. For $T>300$K, there is hopping between the TCs, but this is spatially non-uniform, and even temperature dependent.  We find that the TCs are non-uniformly distributed in space. Volumes with a high concentration of TCs  have longer trap lifetimes than volumes with few or no TCs. The barriers between TCs that are close together tend to be small, enabling rapid hopping {\it within the high density region}, but the effect of a collection of TCs in close proximity is to create a strong barrier for the Ag to escape to another volume. Thus, one can introduce the notion of ``supertraps"  or cages, associated with a high concentration of TCs. Ag dynamics at finite temperature can be understood to be cage-cage hopping with larger trapping time at low temperatures. Our simulations provide dynamics reminiscent of motion in supercooled colloids\cite{weeks02} and diffusion of Li ions in silicate glass\cite{habasaki97,habasaki99,habasaki02}. 

The results reported here are from the plane-wave \emph{ab initio} code VASP\cite{vasp} (Vienna {\it ab initio} simulation package) at constant volume.  We began with a 239 atom (GeSe$_{3}$)$_{0.85}$Ag$_{0.15}$ model (50 Ge, 153 Se and 36 Ag atoms), generated with the  local basis {\it ab initio} FIREBALL method\cite{polarization}, with a neutron structure factor close to experiment. We relaxed this with VASP, and only small changes were noted. Details of coordination and structure are available elsewhere\cite{tafen242}  These models possess the interesting feature that all Ag atoms are found halfway between pairs of host atoms  (Ge-Ag-Ge, Se-Ag-Se or Ge-Ag-Se), with the exception of only one Ag. About 61\% of Ag sites reside between a pair of Se, the rest involve one Ge. The distances between host pair atoms is between $4.7$ to $5.2$ \AA ~ and the bond length of the Ag to the atoms of the pair is in the range $2.4$ - $2.6$ \AA. About 17$\%$ of Ag have 2-fold Se neighboring pair, the rest of the Ag host pairs are under-coordinated.  To verify the existence of these traps in an independent way, we introduced unbonded Ag at a variety of locations in a 64-atom amorphous Se model \cite{zhang01} at T=300K, so that the Ag could ``probe" the energy landscape in an unbiased fashion and without exception, the Ag became trapped between two Se host atoms with distances in the range we indicate above for the ternary glass.

\begin{figure}
\resizebox{90mm}{!}{\includegraphics{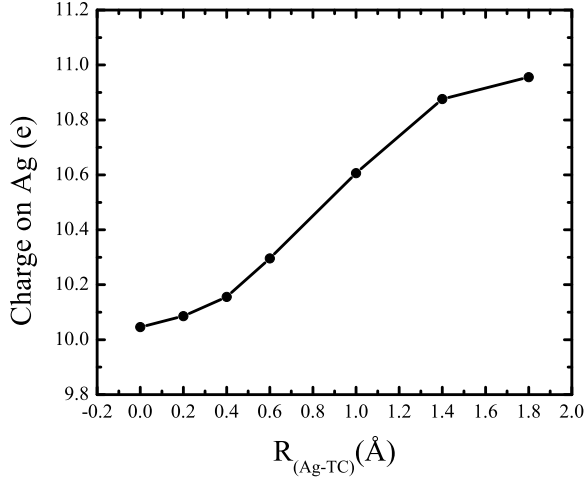}}
\caption{Valence charge on Ag site with respect to the distance from the TC site (see text). The charge state of Ag changes from neutral when isolated to ionic near TC. Neutral silver has 10 3d and one
4s electrons.}
\label{fig1}
\end{figure}

Returning to the ternary material, we begin by reporting some electronic characteristics of the Ag in the solid state.  We examined the charge state of Ag  by removing all but one Ag. It is found that the remaining Ag lost a charge of about one electron when moved from an isolated position to a TC.  Near a TC, Ag acts as a positive ion\cite{kolobov91}. At TC, about 65-70 $\%$ of the charge lost by Ag site is transferred to the neighboring pair sites. Thus, in the Born-Oppenheimer approximation (expected to be valid here), we find Ag to be uncharged while undergoing diffusive motion, and in the Ag$^{+1}$ state for bonded conformations (see Fig. \ref{fig1}). This suggests that the metallic Ag filaments of the PMC are only weakly bonded to the network, while the Ag atoms in the filament are chemically bonded to each other. 

We find that Ag in the network affects the electronic density of states near the optical gap.  For non-bonded Ag, it is found that a silver-related level appears about $0.2$ eV below  the LUMO level. With Ag placed at a TC, an increase in the valence edge level density is apparent, as the charge is transferred from the Ag site to the neighboring sites. Therefore neighbors of Ag are found to contribute to the valence tail and the Ag creates levels near the conduction tail. 

We also added an electron to occupy the LUMO, with Ag placed at TC. It is found that the total charge on Ag sites is increased and consequently the contribution of Ag to the conduction tail is enhanced. The increase of charge on Ag sites resulted in an increase (7-10$\%$) of bondlength between Ag and the neighboring pair sites.  Light of an appropriate wavelength might induce electronic transitions between the valence tail and conduction tail states, both of which are Ag-related. Hellmann-Feynman forces associated with light-induced occupation change may be large for the Ag, and stimulate metal diffusion in a local-heating picture\cite{drabold08}. Also, addition of an electron (mimicking light-induced occupation change) suggests weakened Ag bonding, thus lower barriers, and consequently more diffusive Ag.

\begin{figure}
\resizebox{90mm}{!}{\includegraphics{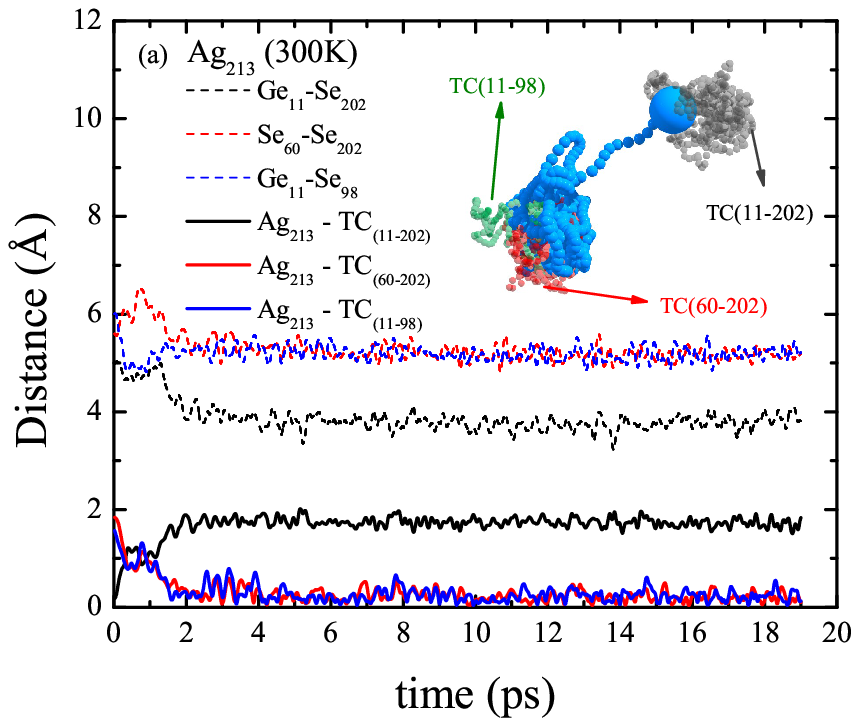}}
\resizebox{90mm}{!}{\includegraphics{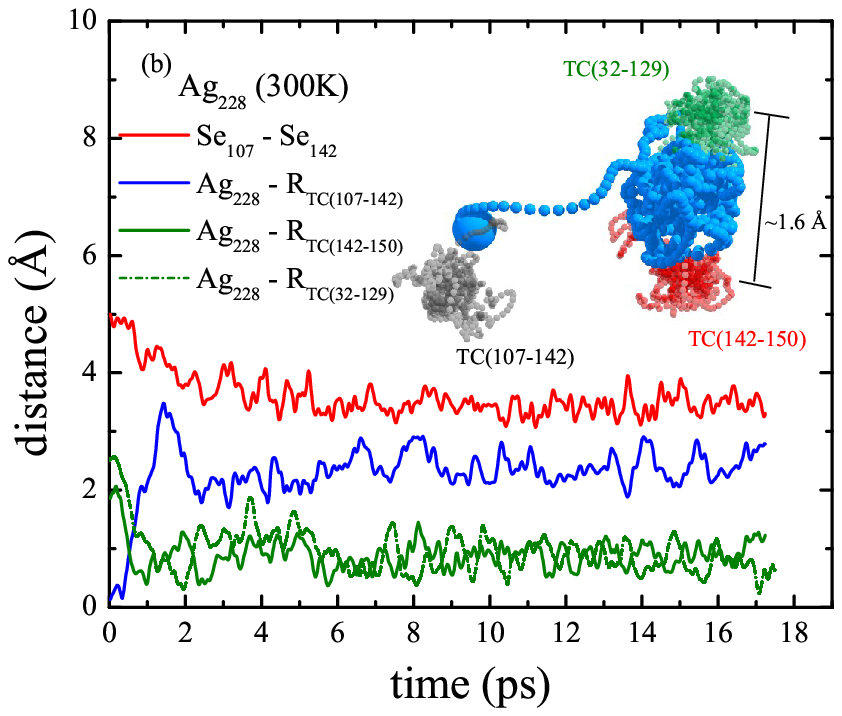}}
  \caption{(color online) Characteristic examples of silver dynamics: Top: ``Type 1" trap, Bottom "Type 2" (see text). The inset shows the trajectories of Ag sites (blue) with the trajectories of neighboring TCs (grey, green and red).}
\label{fig2}
\end{figure}

\begin{figure}
\resizebox{100mm}{!}{\includegraphics{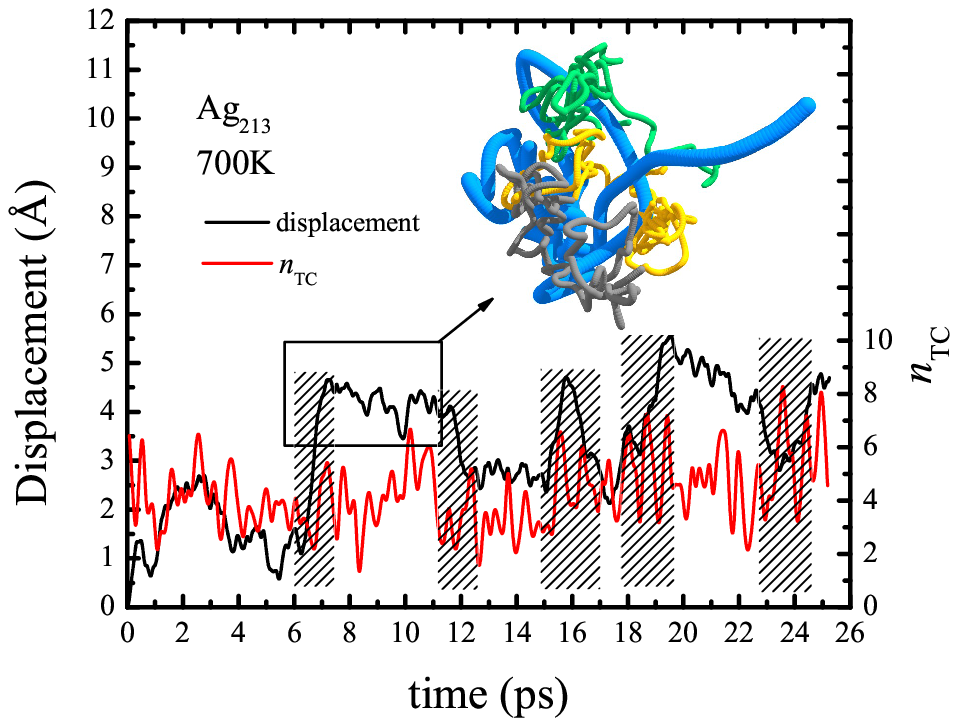}}
\caption{(color online) Displacement of Ag$_{213}$ and the average number $n_{TC}$ of trapping centers surrounding Ag$_{213}$  (within the radius of 4.0 \AA ~ around the Ag atom). The shaded regions highlight the hops. The trajectory of Ag$_{213}$ (blue) is shown along with those of three neighboring TCs (yellow, green and grey) in the time during which Ag is trapped after making a `jump'.}
\label{fig3}
\end{figure}

Silver dynamics were studied by constant-temperature Nos\'e-Hoover dynamics at 300K and 700K. Extended trajectories of 20ps were obtained. At 300K, silver is largely trapped: only 6 hopping events were observed. The silver traps fall into two categories. {\it Type ~1~ (32$\%$)} are  strongly bound: 4 Ag atoms sit at single TC with no neighboring TC within a radius of 2.0 \AA, and 7 Ag occupy two overlapping TCs with the host pairs making an angle of about 90$^{o}$ to each other. {\it Type ~2~ (68$\%$)} are oscillating between two or three closely spaced TCs. Fig. \ref{fig2} reveals the dynamics of the two types of Ag. In the top panel of Fig. \ref{fig2}, dynamics of Ag$_{213}$ (type 1) relative to the three TCs is shown. Initially it is trapped at TC(11-202) (between Ge$_{11}$ and Se$_{202}$), the gradual decrease in the Ge$_{11}$-Se$_{202}$ distance pushes the Ag out and it is eventually trapped at two overlapping TCs [TC(60-202) and TC(11-98)]. Note the stabilization in the Se$_{60}$-Se$_{202}$ and Ge$_{11}$-Se$_{98}$ distances after Ag is trapped between the host atom-pair.  The bottom panel of Fig. \ref{fig2} illustrates type 2 Ag silver motion. Ag$_{228}$ is initially trapped at TC(107-142). It becomes unstable due to the motion of TC(142-150), (initially at 1.8 \AA  ~ from Ag$_{228}$), and then a decrease in the Se$_{107}$-Se$_{142}$ distance moves it out of its initial TC. Eventually Ag$_{228}$ is trapped between the two TCs [TC(142-150) and TC(32-129)] with an average distance of about 1.6\AA ~ between them. The trajectory of Ag$_{228}$ shows cage or ``super-trapping" between two TCs. Note the larger fluctuations in the position of Ag as compared to type 1 Ag (trajectory of Ag$_{213}$). The hopping lengths between one TC to other TCs, in general depend upon the concentration of neighboring TCs. A  larger number of neighboring TCs tends to confine the Ag in a smaller region (1.0 \AA) as in the two cases discussed above, while larger jumps are observed for Ag with lower concentration of neighboring TCs.

\begin{figure*}
\resizebox{140mm}{!}{\includegraphics{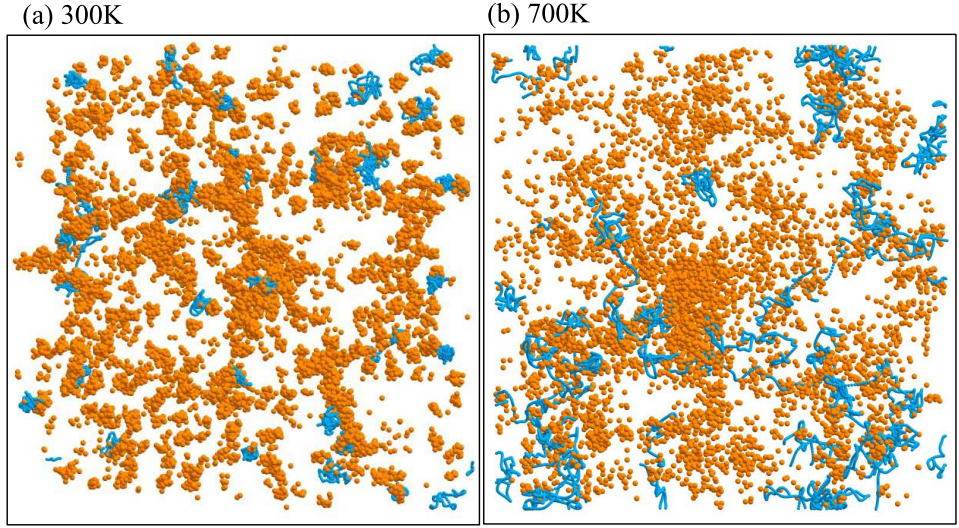}}
\caption{(color online) Trajectories of TCs (orange) and all Ag (blue) in equilibrium (of about 6 ps length) are shown at (a) 300 K and (b) 700K. Concentration of Ag sites around dense regions of TCs is apparent at 300K and the Ag hops between the dense regions of TCs are clear at 700K.}
\label{fig4}
\end{figure*}

The mean squared displacement of Ag at 700 K shows a linear increase with time, illustrating the diffusive nature of the Ag dynamics, consistent with the previous studies \cite{tafen242}. The Ag dynamics consists of a gradual drift away from the initial (fully relaxed) TC configurations,  followed by hops between cages. There are 20 jumps observed, much larger than that for 300K, as expected, with an average time period of {\it ca.} 7 ps between the hops. The hopping lengths vary between 1.5 - 4.0 \AA. We characterize such hopping dynamics in terms of the variation in the concentration of TCs $n_{TC}$ around Ag sites. Fig. \ref{fig3} shows the displacement of Ag$_{213}$ at 700K. The hopping is apparent in the form of abrupt changes in the displacement. To understand these jumps, we counted the number of TCs surrounding the Ag site in a radius of 4.0 \AA. In Fig. \ref{fig3}, the concentration of neighboring TCs is also plotted with the displacement. A correlation between the hops and the decrease in $n_{TC}$ is apparent. The jumps tend to occur at times when either $n_{TC}$ is low or exhibits a sudden decrease. Also, the figure reveals the significant impact of thermal fluctuations on the TCs and their density. The trajectory of Ag$_{213}$ along with the trajectories of three neighboring TCs (inset) in the trapped region gives further insight into the nature of the trap.   At higher temperature, the Ag sites are more unstable because of thermal fluctuations in the neighboring network and their higher thermal energies. It would require a higher density of TCs to confine the Ag dynamics. The hops can be considered as a spontaneous event, which may be triggered by a decrease in the concentration of neighboring TCs. 

Fig.\ref{fig4} illustrates Ag motion at both temperatures. In equilibrium, the Ag at 300K are confined on or in between the dense regions spanned by the TCs. Note that the volume fraction containing no TC is large at 300K. At 700K, TC are less concentrated owing to the thermal fluctuations in the host network, thus enhancing Ag diffusion. The Ag jumps between dense TCs regions (cages) is apparent. Such dynamics is quite similar to the hopping dynamics suggested for Li ions in silicate glasses \cite{habasaki99,habasaki97}, where the high mobility of Li ions is correlated with the decrease in the volume fraction of voids, which decreases the local atomic density around ion \cite{habasaki02,weeks02}. At 300K, the TCs are relatively more stable and are distributed randomly as shown in Fig. \ref{fig4} in the form of dense and dilute regions, similar to the Scher-Lax-phillips traps \cite{phillips96}. One can view Fig. \ref{fig4} as a revealing a percolation-like process: at the higher temperature the trapping basins become more extended and overlapping, until transport through the glass becomes possible.

In summary, we have presented a microscopic picture of Ag dynamics in the (GeSe$_{3})_{75}Ag_{15}$ glass at low and high temperatures. It is shown that the Ag is loosely bonded to the host network and the low energy sites of Ag at 0K are at the centers of the pairs of host sites, having distances of about 5.2 \AA ~ separating them. At low temperatures, the Ag is mainly trapped at or in between such trapping centers. At higher temperatures, the dynamics mainly consist of trapped motion followed by hops. The trapped region of Ag consists of higher concentration of TCs forming a cage like network, which confine Ag dynamics in the cage. Large fluctuations in the concentration of TCs trigger hopping and allow the Ag to move between such cages.

We gratefully acknowledge the National Science Foundation for support under grants DMR-0600073 and DMR-0605890. DAD thanks the Leverhulme Trust (UK) and the NSF International Materials Institute for New Functionalities in Glass under award DMR-0409588 for supporting his sabbatical visit to the University of Cambridge. We thank Prof. S. R. Elliott and Dr. Jim Phillips for many helpful discussions, and thank Trinity College, Cambridge for hospitality.


\begin{thebibliography}{100}
\bibitem{polarization} FIREBALL did not include polarization orbitals; VASP uses an essentially complete basis.
\end{thebibliography}
\end{document}